\newcommand{\df}{{\rm d}}

\newcommand{\gev}{\ {\rm GeV}}

\newcommand{\zp}{Z.\ Phys.\ }
\newcommand{\pl}{Phys.\ Lett.\ }

\newcommand{\be}{\begin{equation}}
\newcommand{\ee}{\end{equation}}
\def\extps#1#2#3#4#5#6#7{
	\vbox{\vskip#7 \hbox{
		\special{ps: plotfile #1 asis}
		\special{ps::[asis,end]
		restore
		0 SPE
		}
		\hskip#6 }
		}
	}

\documentstyle[12pt]{article}

\newlength{\dinwidth}
\newlength{\dinmargin}
\begin{document}
\title{
\hfill {\large DTP--93--42}\\
 \vspace{1 cm}
        {\bf Coherence effects in the current fragmentation region at HERA}
        \thanks{To appear in Proceedings of the Workshop "Hera--the New
Frontier
        for QCD", Durham, March 1993.}
\author{
    K.Charchu\l a
    \thanks{permanent address: Dept. of Physics, Warsaw University,
     Hoza~69, PL--00--681 Warsaw, Poland}
\\[10pt]
    Department of Physics\\
      University of Durham\\
    Durham DH1 3LE, UK\\[15pt]
       }
\date{ }
     }

\maketitle

\begin{abstract}
The structure of  the hadronic final state in deep inelastic scattering
at HERA is studied on the level of the Monte Carlo event generator.
Special emphasis is given to the colour coherence phenomenon in
the current fragmentation region. It is shown that results of perturbative
QCD recently tested in $e^+e^-$ annihilation at LEP also describe  production
of hadrons in current fragmentation region of deep inelastic scattering.
\end{abstract}

\setcounter{page}{1}
\vspace{0.5 cm}

\section{\bf Introduction}

Study of the hadronic final state in deep inelastic scattering (DIS) provides
us  with detailed information about the space--time picture in  the $ep$
process.
There are three main phases in the picture:
partons taking part in hard scattering emerge as well defined jets,
evolution of parton cascade gives rise to the bulk of rather soft
radiation which populates the original jet with secondary partons and finally
the development of  parton system  ends in a hadronization stage when
coloured partons are transformed into hadrons.

Particles produced in the DIS process which is usually characterized  by
$(x_{Bj}, Q)$ variables can be classified as two distinct regions.
Those coming from the
evolution of the struck (current) quark belong to the current fragmentation
(CF) region while particles produced from parton radiation in the
initial state form the target fragmentation (TF) region. Evolution of
the parton system is quite different in these two regions.
While  $Q$ scale is in principle the only parameter
which determines inclusive particle distributions in the CF region,
hadronic spectra in the TF region depend essentially upon both
$x_{Bj}$ and $Q$   variables.
In order to separate these two regions kinematically for subsequent independent
study one usually discusses the  distributions of final particles
in Breit frame where the virtual photon with 4-momentum
$Q=(0,-2x_{Bj}{\bf P}$) moves antiparallel to the target (proton) with
momentum
${\bf P}$.
In this frame  products of the
fragmentation of the target and the current jet move in  opposite directions
(at least in the  simple Quark Parton Model picture).

The final state properties of hadrons in CF region are expected to be similar
to those of hadrons produced in $e^+e^-$ annihilation. On the other hand
the internal structure of the TF region is much more complicated \cite{gri88}.
 From the analysis of hadronic events in $e^+e^-$ it is known that the
phenomenon of coherence of soft gluon emissions modifies the QCD
bremsstrahlung of partons in a timelike region (CF region
of DIS) in an essential way (for review, see \cite{dok91}).
This effect was also predicted to hold in the spacelike cascade, i.e. TF region
of DIS.  In both cases coherence leads to angular ordering of  sequential
parton decays in the parton cascade which manifests in the depletion of
emission of soft particles.
The first experimental analysis of coherence effects in DIS
done by EMC \cite{emc87} showed promising results.

HERA $ep$ collider with its vast region of $Q$ (up to $\sim 300 \gev$) and
$x_{Bj}$ (down to $\sim 10^{-5}$) variables offers new experimental
possibilities for the investigation of  the structure of the hadronic state.
 Study of particle production in the target fragmentation region will be a
particularly challenging task. But  analysis of  particle spectra in
the current fragmentation region can provide also  better insight into the way
the timelike cascade is developing and the hadronization phase is set up.

In this contribution I present a Monte Carlo (MC) study of some inclusive
distributions of particles in the CF region of DIS at HERA
$ep$ collider and compare them with
analytical results of perturbative QCD (pQCD) which were recently confirmed by
experiments at $e^+e^-$ LEP collider \cite{dok91,dok92}. This study should
serve as an  indication of what we can learn from looking at inclusive
spectra in DIS rather than a  rigorous analysis \cite{cha93}.
In section 2 I briefly review the phenomenological situation relevant for
this study.
The comparison of MC results in CF region with  pQCD predictions is presented
in section 3. Finally,  section 4 contains summary of   the results.

\section{\bf Coherence effects in time-like cascades}

Recent results from LEP experiments have demonstrated the QCD bremsstrahlung
nature of hadroproduction.
They have also confirmed several pQCD predictions
for global features of the hadronic system, such as the inclusive
momentum spectra, the mean multiplicities, etc.\cite{dok92}.
 In the following by pQCD
results I will mean results obtained in the Modified Leading Logarithmic
Approximation (MLLA) accompanied by the hypothesis of Local Hadron Parton
Duality (LPHD) \cite{dok91}. This hypothesis allows the results obtained on the
parton level to be applied in the analysis of experimental data (hadron level).

The important role in the multihadron production is played by QCD coherence.
The basic consequence of this phenomenon (also called
destructive interference) is angular ordering--a decrease of successive
opening angles in the parton cascade. As a result a decrease in soft
particle production is observed. Inclusive distributions of particles were
found to be sensitive to the soft gluon coherence. In particular, coherence
leads to a slower rise in the gluon multiplicity within a parton jet
and to change in the shape of the gluon momentum distribution (relative
to the situation without coherence). These effects were observed experimentally
in $e^+e^-$ annihilation.

The inclusive momentum distribution in the CF region of DIS can be related
to that of $e^+e^-$ annihilation:
\be
\frac{1}{\sigma}\frac{\df\sigma^{\rm CF}}{\df\ln(1/x_p)} =
\frac{1}{2\sigma}\frac{\df\sigma^{e^+e^-}}{\df\ln(1/x_p)}
\ee
where $x_p = 2p_h/\sqrt{s}$ and  $\sqrt{s} \stackrel{\rm DIS}{=} Q$.
It was found at LEP \cite{opa90} that already a gaussian  approximation to the
exact MLLA result \cite{dok91} describes the shape of momentum distribution
quite well
(especially  around peak position):
\be
\label{gauss}
\frac{1}{\sigma}\frac{d\sigma^{e^+e^-}}{{\rm d}\ln(1/x_p)} =
{\cal N}(Y) \left(\frac{36N_C}{\pi^2 b Y^3}\right)^{1/4}
{\rm exp}
\left[ - \sqrt{\frac{36N_C}{b}}\frac{(l - \ln(1/x_{max}))^2}{Y^{3/2}}\right]
\ee
with $l = \ln(1/x_p)$, $Y = \ln(\sqrt{s}/2\Lambda_{eff})$ and
$b = \frac{11}{3}N_c - \frac{2}{3}N_f$.
At  large $Y$ the average multiplicity $\cal N$ has the following
asymptotic behaviour:
\be
\label{mult}
{\cal N}(Y) = K^{ch} Y^{1/4 - B/2} \exp{\sqrt{\frac{16N_c}{b} Y}}
\ee
where $B = a/b$ and $a = \frac{11}{3}N_c + \frac{2}{3} N_f/N_c^2$.
Finally, the position of the maximum of the distribution is also predicted by
the theory:
\be
\label{peak}
\ln(1/x_{max}) = 0.5Y + B\sqrt{\frac{b}{16N_C}}\sqrt{Y} + {\cal O}(1)
\ee
where the ${\cal O}(1)$ term containing higher order corrections is  given
by pQCD.
In the case of charged particle spectrum, the only two free parameters
appearing
in the pQCD formulas were found by OPAL  \cite{opa90} to be:
$\Lambda_{eff} \simeq 0.203 \gev$ and $K^{ch}\simeq 0.19$. In addition, the
size of
higher  order corrections was estimated in that analysis
 (${\cal O}(1) \simeq - 0.38$) to be in agreement with theory predictions
 \cite{dok92}.

Perturbative QCD not only predicts the shape of  gluon momentum distribution
at fixed energy $\sqrt{s}$ but also describes the energy evolution of the
spectrum, eq.(\ref{peak}). In fact, since hadronization effects should be
independent
of  CM energy of hard scattering, analysis of the evolution of  hadron
spectrum with $s$ should provide direct information on the energy evolution of
the underlying gluon spectrum. In this case, study of CF region in DIS gives a
whole
range of $\sqrt{s} = Q$ values  ($Q\sim 2-300 \gev$) in one experiment, in
contrary to
$e^+e^-$ experiments where one (LEP) or  a few (TASSO) energies were
possible.

A third observable which shows influence of coherence effects and which
in some sense provides complementary information to
that of the evolution of the peak is
average multiplicity of produced hadrons.
Again, detailed analysis of this quantity in $e^+e^-$ annihilation was done
at LEP. It was found (see, for example, analysis by DELPHI Coll.\cite{del91})
that  data is very well
described by the following formula obtained in MLLA approximation
\cite{dok91,web84}:
\be
\label{webber}
<n_{ch}(s)> = a\alpha_s(s)^{b_1} \exp[b_2/\sqrt{\alpha_s(s)}]
\ee
with constants $b_1 = 0.49$, $b_2 = 2.26$ given by theory,
$\alpha_s$ calculated from the 2--loop formula and $a = 0.067$,
$\Lambda = 0.145\gev$ fitted to $e^+e^-$ data.
Equivalently, one could also take the  formula (\ref{mult}) with parameters
fixed from the gaussian spectrum (\ref{gauss}).

\section{\bf Inclusive distributions in the CF region of DIS}

\begin{figure}
\hfil\extps{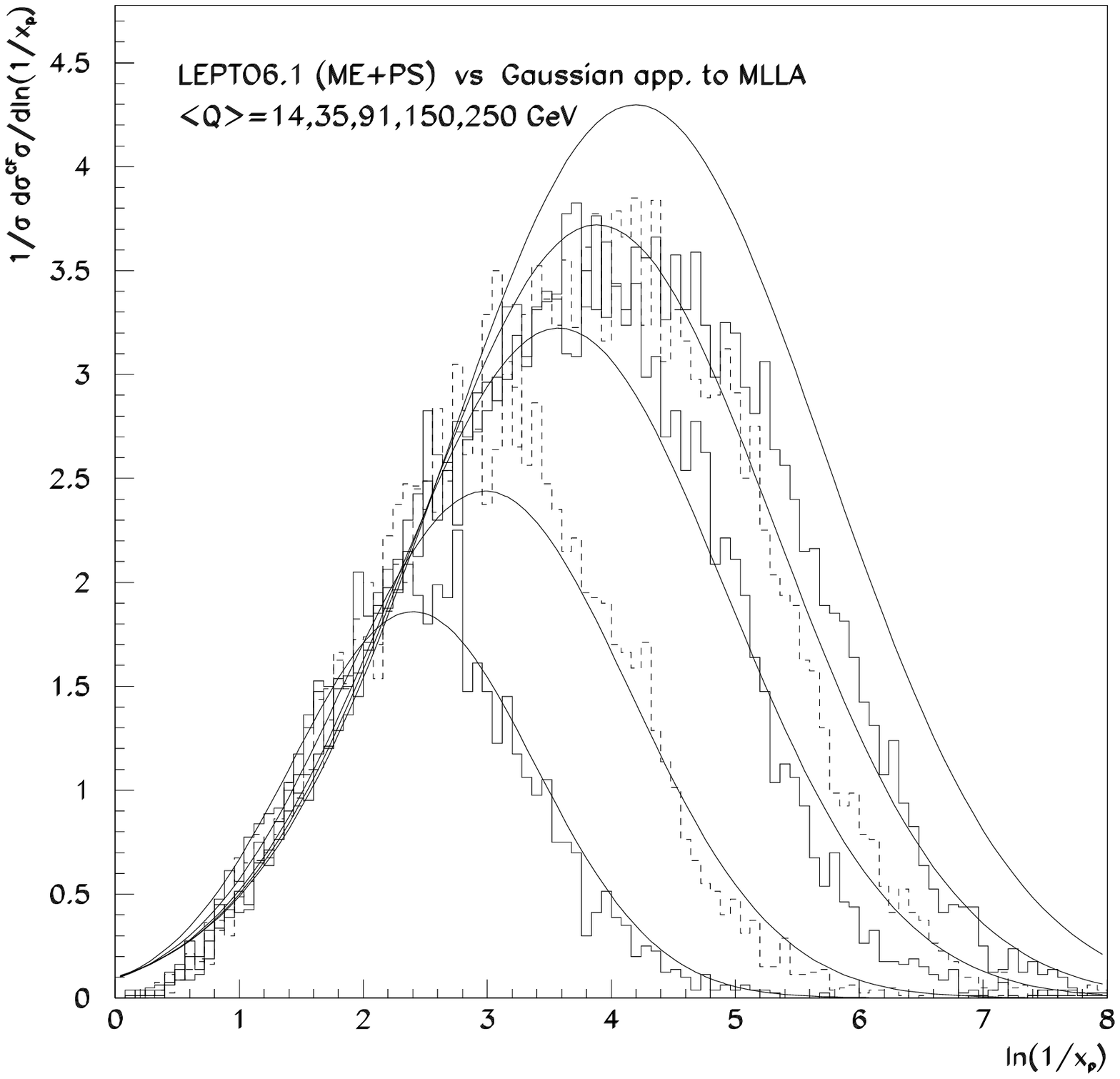}{505 pt}{505 pt}{25 pt}{120 pt}{300 pt}{300 pt}\hfil
\vspace{.25in}
\caption{Momentum distribution of charged particles in CF region as predicted
by
 gaussian approximation (\ref{gauss}) and generated by the ME+PS option
of LEPTO6.1.}
\end{figure}
In this section I compare predictions of pQCD for the above discussed inclusive
particle distributions in the CF region of DIS with  MC generated data in
HERA kinematic region. Free parameters of theoretical formulas were fixed
at $\sqrt{s} = 91\gev$ from $e^+e^-$ LEP experiments shortly discussed
in the previous section. To generate particle distributions,
LEPTO6.1 \cite{lep91} Monte Carlo event generator was used.
The program allows for  the production of a parton system in different
scenarios:
\begin{itemize}
\item
 lowest order hard  scattering (order $\alpha_s^0$) followed by parton shower
with chosen maximum virtuality scale (e.g. PS(Q$^2$), PS(W$^2$) options),
\item
hard scattering to order $\alpha_s$ followed by  parton shower with maximum
virtuality
scale determined by the hard emitted parton (ME+PS option),
\item
order $\alpha_s$ hard scattering without parton cascade phase
(i.e. maximum $2$ partons + proton remnant on the parton level) (ME option).
\end{itemize}
The transformation from parton--to--hadron level is done by the string
fragmentation model (JETSET7.3 \cite{jet92}) which was found to describe
hadroproduction in $e^+e^-$ at LEP very successfully.
Parton shower prescription used in LEPTO program has built in angular ordering
of  subsequent parton emissions both in the development of spacelike and
timelike  cascades.
It was recently demonstrated by HERA experiments \cite{her93}
that ME+PS option of LEPTO gives a rather good
description of global hadronic properties like rapidity distribution,
transverse energy spectrum, etc.

\begin{figure}
\hfil\extps{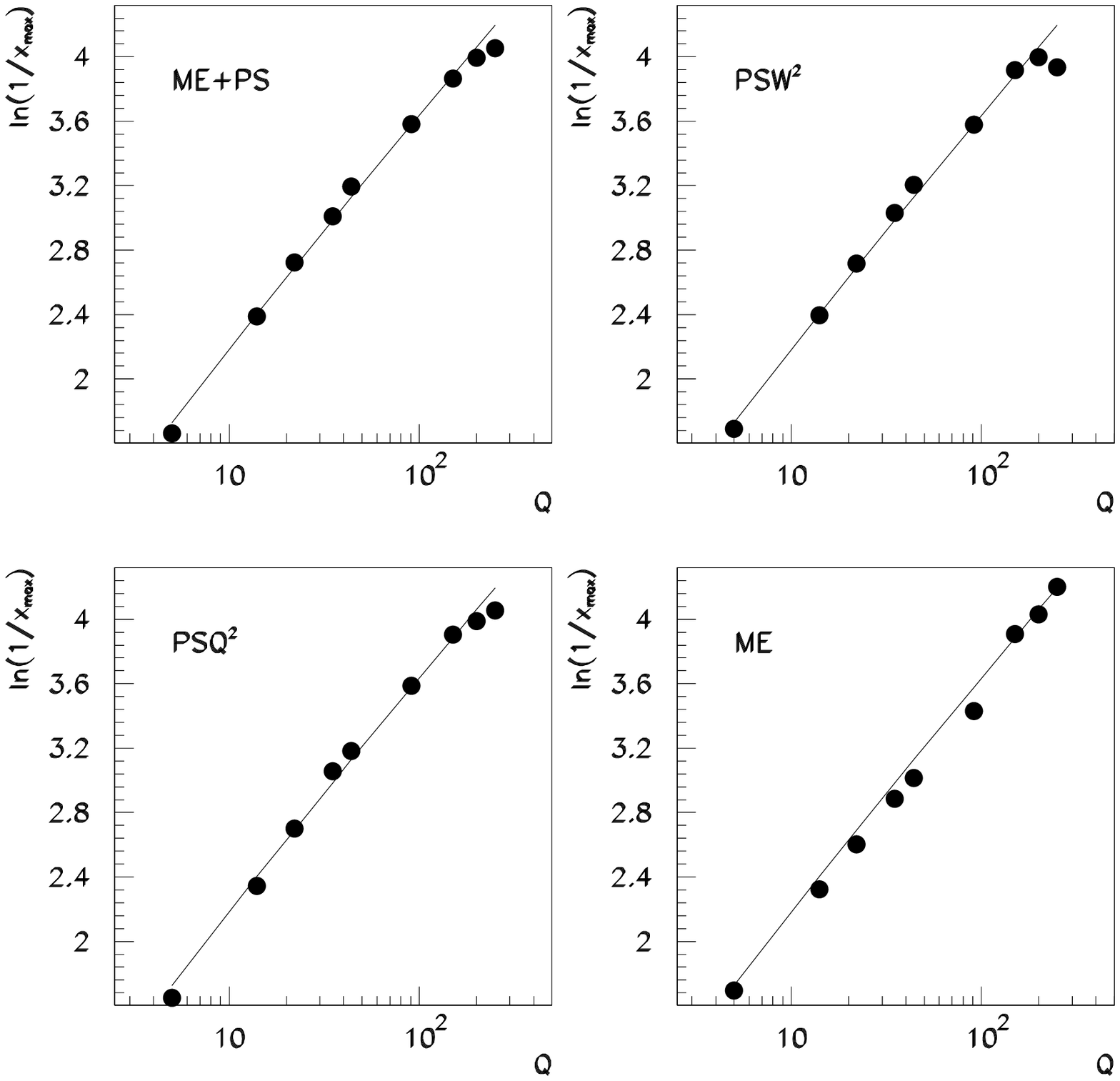}{505 pt}{505 pt}{25 pt}{120 pt}{300 pt}{300 pt}\hfil
\vspace{.25in}
\caption{ Energy evolution of the maximum of $\ln(1/x_p)$ compared
to the analytic formula (\ref{peak})-solid line
 and  various options of LEPTO6.1-full points.}
\end{figure}

Our expectations are that pQCD formulas which so well described inclusive
distributions of hadrons coming from $e^+e^-$ annihilation should also be
applicable to the description of hadronic spectra in CF region of DIS
analysed in the Breit frame.
Here we took a rather conservative approach and studied hadronic spectra
at large $Q$ where  current jet is a well defined object \cite{her93} and
hadroproduction in the CF region is to a good approximation independent
of $x_{Bj}$ value \cite{cha93}.

The shapes of momentum distribution of charged particles at various
$\sqrt{s} = Q$
energies, including $\sqrt{s}= 91\gev$ of $Z^0$ peak, are shown in fig.1.
The lines correspond to the prediction of formula (\ref{gauss}) and histograms
were generated in the ME+PS scenario. It is seen that both the shape and
the position of the maximum produced by LEPTO agree quite well with
predictions of pQCD.
There seems to be a difference in the height of the distribution as produced
by LEPTO and predicted by (\ref{gauss}), especially at
large $Q$. This translates   into differences in  average multplicity which
I shall discuss later. The similar behaviour is also found for PS options of
LEPTO.  On the other hand,  the shape of the distribution produced within
the ME scenario  (without the parton shower phase) badly reproduces  pQCD
shape.

The evolution of the peak position $\ln(1/x_{max})$ (\ref{peak}) of
the momentum distribution with energy $\sqrt{s}$ for various LEPTO options
is shown in fig.2.
The ME+PS (and to a lesser extent also PS)  prescription agrees with
MLLA results (\ref{peak}) quite well.
What is somehow surpising is that  the ME scenario  also gives
similar results in this case despite the fact that the parton cascade
phase is now absent. The only explanation for this is that the string
fragmentation model used for hadronization is able to mimic
to some extent the soft gluon radiation of QCD \cite{azi85}
\footnote{Thanks to V.Khoze for clarifying discussion concerning this point.}.

\begin{figure}
\hfil\extps{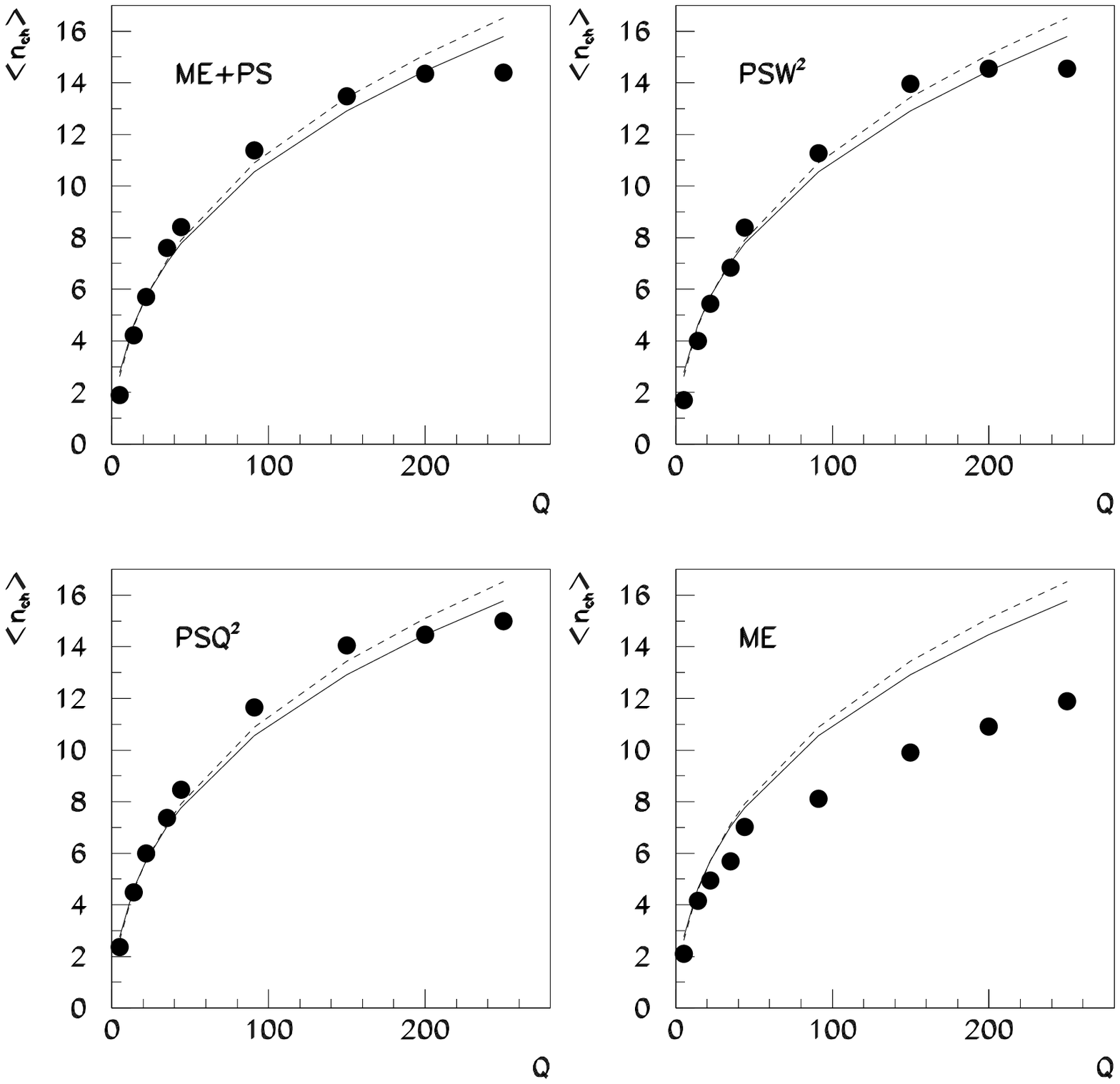}{505 pt}{505 pt}{25 pt}{120 pt}{300 pt}{300 pt}\hfil
\vspace{.25in}
\caption{ Average multiplicity of charged particles compared to
the MLLA results: formula (\ref{webber})-solid line,
  (\ref{mult})-dashed one, and various options of LEPTO6.1.}
\end{figure}

The observable which fully shows  the importance of the parton cascade
phase  is the  average multiplicity of produced hadrons. In fig.3  the average
multiplicity  of charged particles is shown.
Here the solid line corresponds to formula (\ref{webber}), the dashed one
follows from (\ref{mult}) and full circles come from various options of
LEPTO6.1.
It is seen that all the PS options show quite good agreement with
the theory while the plain ME prescription where
hard produced partons directly enter the hadronization phase
fails to produce enough particles in most of  the HERA   $Q$ range.

The above comparisons support our expectations that hadroproduction in CF
region of DIS should manifest the same kind of properties to those  of
$e^+e^-$ annihilation.
It also confirms the significance of the parton cascade phase in the
development of
hadronic final state.
Comparison of analytic predictions of pQCD with  Monte Carlo simulation
points out  several interesting properties of this specific MC program.
It seems that  the choice of a maximum virtuality scale in parton
cascade (two extreme examples of which are $W^2$ and $Q^2$) hardly
shows up in hadron distributions studied in the CF region, at least at large
$Q$ values.
Another observation is that the parton system already with one hard gluon
emitted followed by the string fragmentation model is able to mimic some of the
properties expected from multiple, soft gluon radiation in QCD (i.e. evolution
of the peak of momentum distribution).
These  conclusions may be spoiled at small $Q$ values (corresponding to small
$x_{Bj}$ at HERA) by two effects:
the current jet is no longer a well defined object in this region \cite{her93}
and hadronic spectra in the CF sector show some dependence on $x_{Bj}$ value.

To finalise, let me make  a comment about the other region of particle
production in DIS.
Coherence phenomenon  affects the structure of the parton system
in the TF region in a crucial way, leading to new  effects
which show up most clearly at small $x_{Bj}$ values\cite{gri88}
(for  recent studies of  hadroproduction in the TF region see
 \cite{azi93,ing93,cha93}).
In particular, the shape of inclusive momentum distribution varies both
with $Q$ and $x_{Bj}$   and differs from the gaussian--like form
found in the CF  region.

\section{\bf Summary}

The  analysis of the two distinct regions of DIS
(current and target fragmentation region) in  Breit frame provides us with
rich information about the parton branching phase in both  spacelike and
timelike
cascades, and also about transformation of partons into hadrons (hadronization
phase). In particular, it was shown that coherence effects  show up in the
CF region of DIS in the similar way to that found in  hadroproduction in
$e^+e^-$ annihilation.
HERA kinematic range allows to extend  study of coherence phenomena
in the timelike region far above LEP $\sqrt{s}=91\gev$ energy
(up to $\sqrt{s}\sim 300\gev$)
and for the first time permits testing of pQCD ideas about partonic structure
of
the  spacelike cascade.

\noindent Finally, there are several questions/problems which this initial
study did not address \cite{cha93}:
\begin{itemize}
\item
correlation between the current and target fragmentation regions, particularly
at small $x_{Bj}$ values;
\item
description of hadroproduction in the CF region of DIS by  other existing
MC generators (especially HERWIG \cite{her88} where the
approach closely follows  that of MLLA and LPHD); comparison of pQCD
predictions
for the TF region with MC results \cite{azi93,ing93};
\item
experimental separation of the CF and TF regions in HERA conditions,
transformation to Breit frame, identification of different particle species
in these two regions.
\end{itemize}

\vspace{1cm}
\noindent {\Large\bf Acknowledgments}

I am indebted to  V.Khoze and J.Newton for reading  the manuscript and remarks.
I would also like to thank the organizers  of the Durham Workshop for a
stimulating and enjoyable meeting.
I am   grateful to the Physics Department and Grey College of the
University  of Durham for their warm hospitality.
This work has been partially supported   by the EC "Go--West" Scholarship.




\end{document}